# Fragment Descriptors in Virtual Screening

Igor I. Baskin

Department of Chemistry, Lomonosov Moscow State University, Moscow 119991, Russia

Alexandre Varnek*

Laboratoire d'Infochimie, UMR 7177 CNRS, Université de Strasbourg,
4, rue B. Pascal, Strasbourg 67000, France,
e-mail : *varnek@chimie.u-strasbg.fr*, URL : *http://infochim.u-strasbg.fr*

**Abstract**. This article reviews the application of fragment descriptors at different stages of virtual screening: filtering, similarity search, and direct activity assessment using QSAR/QSPR models. Several case studies are considered. It is demonstrated that the power of fragment descriptors stems from their universality, very high computational efficiency, simplicity of interpretation and versatility.

1. **Introduction**

Chemogenomics aims to discover active and/or selective ligands for biologically related targets by conducting screening, ideally, of all possible compounds against all possible targets, or at least, in practice, available libraries of compounds against main target families [1]. One can hardly imagine to screen experimentally the chemical universe containing from $10^{12}$ to $10^{180}$ druglike compounds [2] against biological target universe. While the number of compounds that can be screened in the course of high-throughput screening experiments can hardly be imagined to exceed several millions per biological target even in the largest pharmaceutical corporations, the number of molecules that can be screened *in silico* in a single non-expensive computational study currently reaches $10^{12}$, and this number can grow sharply in the nearest future. Therefore, the virtual, or *in silico*, screening approaches play a key role in chemogenomics.

Virtual screening is usually defined as a process in which large libraries of compounds are automatically evaluated using computational techniques [3]. Its goal is to discover putative hits in large databases of chemical compounds (usually ligands for biological targets) and remove molecules predicted to be toxic or those possessing unfavorable pharmacodynamic or pharmacokinetic properties. Generally, two types of virtual screening are known: structure-based and ligand-based. The former explicitly uses 3D structure of a biological target at the stage of hit detection, whereas the latter uses only information about structure of small molecules and their properties (activities), see [4-6] and references therein. Structure-based virtual screening (docking, 3D pharmacophores) has been described in series of review articles, see [7-9] and references therein.

In this paper mostly ligand-based virtual screening involving fragment descriptors is considered. Fragment descriptors, represent selected substructures (fragments) of 2D molecular graphs and their occurrences in molecules; they constitute one of the most important types of molecular descriptors [10]). Their main advantage is related to simplicity of their calculation, storage and interpretation (see review articles [11-15] and book chapter [16]). From the mathematical point of view, they constitute a basis of invariants of labeled molecular graphs, and therefore any molecular property and any similarity measure for molecular graphs can be expressed through them [17-19]. Fragment descriptors are information-based ones [20] which tend to code the information stored in molecular structures. This contrasts with knowledge-based (or semiempirical) descriptors issued from the consideration of the mechanism of action. Selected descriptors form a "chemical space" in which each molecule is represented as a vector [21]. Due to their versatility, fragment descriptors could be efficiently used to create a chemical space which separates active and non-active compounds.

Historically, molecular fragments were used in first additive schemes developed in 1950-ies to estimate physicochemical properties of organic compounds by Tatevskii [22, 23], Bernstein [24], Laidler [25], Benson and Buss [26] and others. The Free-Wilson method [27], one of the first QSAR approaches invented in 1960-ies, is based on the assumption of the additivity of contributions of structural fragments to the biological activity of the whole molecule. Later on, fragment descriptors were successfully used in expert systems able to classify chemical compounds as active or inactive with respect to certain type of biological activity. Hiller [28, 29], Golender and Rosenblit [30, 31], Piruzyan, Avidon *et al* [32, 33], Cramer [34], Brugger, Stuper and Jurs [35, 36], and Hodes *et al* [37] pioneered in this field.

An important class of fragmental descriptors, so-called *screens* (structural *keys*, *fingerprints*), has been developed in seventies [38-42]. As a rule, they represent bit strings which can effectively be stored and processed by computers. Although their primary role is to provide efficient substructure searching capabilities in large chemical databases, they are efficiently used for similarity searching [43, 44], to cluster large data sets [45, 46], to assess chemical diversity [47], as well as to conduct SAR [48] and QSAR [49] studies. Nowadays, application of modern machine-learning techniques significantly improves predictive performance of structure-property models based on fragment descriptors [50].

This paper briefly reviews the application of fragment descriptors in virtual screening of large libraries of organic compounds focusing mostly on its three approaches: (i) filtering, (ii) similarity search, and (iii) direct activity/property assessment using QSAR/QSPR models.

**2. Types of fragment descriptors**

Due to their enormous diversity, one could hardly review all types of 2D fragment descriptors used for structural search in chemical database or in SAR/QSAR/QSPR studies. Here, we focus on some of them which are the most efficiently used in virtual screening and *in silico* design of organic compounds.

Generally, molecular fragments can be classified with respect to their topology (atom-based, chains, cycles, polycycles, etc), information content of vertices in molecular graphs (atoms, groups of atoms, pharmacophores, descriptor centers) and the level of abstraction when some information concerning atom and bond types is omitted.

Purely structural fragments are used as descriptors in ACD/Labs [51], NASAWIN [52], ISIDA [15] and some other programs. These are 2D subgraphs in which all atoms and/or bonds are represented explicitly and no information about their properties is used. Their typical examples are the sequences of atoms and/or bonds of variable length, branch fragments, saturated and aromatic cycles (polycycles) and atom-centered fragments (ACF). The latter consist of a single central atom surrounded by one or several shells of atoms with the same topological distance from the central one. The ACF were invented by Tatevskii [22] and Benson and Buss [26] in 1950-ies as elements of additive schemes for predicting physicochemical properties of organic compounds. In earlier seventies, Adamson [53] investigated the distribution of one shell ACF in some chemical databases with respect to their possible application as screens. Hodes reinvented one shell ACF as descriptors in SAR studies under the name *augmented atoms* [37], and also suggested *ganglia augmented atoms* [54] representing two shells ACF with generalized second-shell atoms. Later on, one shell ACF were implemented by Baskin *et al* in the NASAWIN [52] software and by Varnek *et al* in ISIDA [15, 55] package. Atom-centered fragments with arbitrary number of shells were implemented by Poroikov *et al* in the PASS [56] program under the name *multilevel neighborhoods of atoms* [57], by Xing and Glen under the name *tree structured fingerprints* [58] (sometimes referred to by Bender, Glen et al as *atom environments* [59, 60] and *circular fingerprints* [61-63]), and by Faulon under the name *molecular signatures* [64-66].

It has been found that characterizing atoms only by element types is too specific for similarity searching and therefore doesn't provide sufficient flexibility needed for large-scaled virtual screening. For that reason, numerous studies were devoted to increase an informational content of fragment descriptors by adding some useful empirical information and/or by representing a part of molecular graph implicitly. The simplest representatives of those descriptors were *atom pairs and topological multiplets* based on the notion of *descriptor center* representing an atom or a group of atoms which could serve as centers of intermolecular interactions. Usually, descriptor centers include heteroatoms, unsaturated bonds and aromatic cycles. An *atom pair* is defined as a pair of atoms (**AT**) or descriptor centers separated by a fixed topological distance: **$AT_i$-$AT_j$-*Dist*,** where $Dist_{ij}$ is the shortest path (the number of bonds) between $AT_i$ and $AT_j$. Analogously, a *topological multiplet* is defined as a multiplet (usually triplet) of descriptor centers and topological distances between each pair of them. In most of cases, these descriptors are used in binary form in order to indicate the presence or absence of the corresponding features in studied chemical structures.

The atom pairs were first suggested for SAR studies by Avidon under the name *SSFN* (*Substructure Superposition Fragment Notation*) [33, 67]. Then they were independently reinvented by Carhart and co-authors [68] for similarity and trend vector analysis. In contrast to SSFN, Carhart's atom pairs are not necessarily composed only of descriptor centers, but account for the information about element type, the number of bonded non-hydrogen neighbors and the number of π electrons. Nowadays, Carhart's atom pairs are rather popular for conducting virtual screening. Modified Carharts's atom pairs, in which descriptors centers are represented by atoms coded with the help of a hierarchical classification scheme [52], have recently been used in virtual screening based on the one-class classification technique [69]. *Topological Fuzzy Bipolar Pharmacophore Autocorrelograms* (*TFBPA*) [70] by Horvath are based on atom pairs, in which real atoms are replaced by pharmacophore sites (hydrophobic, aromatic, hydrogen bond acceptor, hydrogen bond donor, cation, anion), while $Dist_{ij}$ corresponds to different ranges of topological distances between pharmacophores. These descriptors were successfully applied in virtual screening against a panel of 42 biological targets using similarity search based on several fuzzy and non-fuzzy metrics [71], performing only slightly less well than their 3D counterparts [70].

*Fuzzy Pharmacophore Triplets* (*FPT*) by Horvath [72] is an extention of (*Fuzzy Bipolar Pharmacophore Fingerprints*) *FBPF* [71] for three sites pharmacophores. An important innovation in the *FPT* concerns accounting for proteolytic equilibrium as a function of pH [72]. Due to this feature, even small structural modifications leading to a $pK_a$ shift, may have a profound effect on the fuzzy pharmocophore triples. As a result, these descriptors efficiently discriminate structurally similar compounds exhibiting significantly different activities [72].

Some other topological triplets should be mentioned. Thus, *Similog pharmacophoric keys* by Jacoby [73] represent triplets of binary coded types of atoms (pharmacophoric centers) and topological distances between them. Atomic types are generalized by four features (represented as four bits per atom): potential hydrogen bond donor or acceptor; bulkiness and electropositivity. The *topological pharmacophore-point triangles* implemented in the MOE software [74] represent triplets of MOE atom types separated by binned topological distances. Structure-property models obtained by support vector machine method with these descriptors have been successfully used for virtual screening of COX-2 inhibitors [75] and $D_3$ dopamine receptor ligands [76].

*Topological torsions* by Nilakantan *et al* [77] is a sequence of four consecutively bonded atoms $AT_i$-$AT_j$-$AT_k$-$AT_l$, where each atom is characterized by a number of parameters similarly to atoms in Carhart's pairs. In order to enhance efficiency of virtual screening, Kearsley *et al* [78] suggested to assign atoms in the Carhart's atom pairs and Nilakantan's topological torsions to one of seven classes: cations, anions, neutral hydrogen bond donors, neutral hydrogen bond acceptors, polar atoms, hydrophobic atoms and other.

In the frameworks of the Simplex representation of molecular structure (SiRMS) by Kuz'min *et al* [79] any molecule can be represented as an ensemble of different 4-atomic fragments - simplexes. The occurrence number of identical simplexes in a molecule is a descriptor value. Subgraphs corresponding to simplexes can be either connected or disconnected. Not only atom type, but also some other physical-chemical characteristics of atoms (partial charge, lipophilicity, refraction and atom's ability for being a donor/acceptor in hydrogen-bond formation, etc) can be used for atom labeling. Fragment descriptors based on SiRMS were used for building QSAR models and performing virtual screening and molecular design of compounds with strong antiviral activity [80].

ISIDA Property-Labeled Fragment Descriptors (IPLF) were introduced as counts of specific subgraphs in which atom vertices are colored with respect to some local property [81]. Various coloring strategies (notably pH-dependent pharmacophore and electrostatic potential-based flagging) can be combined in the framework of this approach with various fragmentation schemes (chains, atom pairs, augmented atoms, trees, etc). IPLF showed excellent results in similarity-based virtual screening for analogue protease inhibitors and in building highly predictive octanol-water partition coefficient and hERG channel inhibition models [81]. IPLF were also advantageously used in combination with the Generative Topographic Mapping (GTM) method for chemical space visualization, structure-activity modeling and database comparison [82].

In contrast to QSPR studies based mostly on the use of complete (containing all atoms) or hydrogen-suppressed molecular graphs, handling biological activity at the qualitative level, often demands more abstractions. Namely, it is rather convenient to approximate chemical structures by *reduced graphs*, in which each vertex is an atom or a group of atoms (descriptor or pharmacophoric center), whereas each edge is a topological distance $Dist_{ij}$. Such biology-oriented representation of chemical structures was suggested by Avidon *et al* as descriptor center connection graphs [33]. Gillet, Willett and Bradshaw have proposed the GWB-reduced graphs which use the hierarchical organization of vertex labels. This allows one to control the level of their generalization which may explain their high efficiency in similarity searching.

## 3. Application of fragment descriptors in virtual screening and in silico design

In this chapter application of fragment descriptors at different stages of virtual screening is considered.

### 3.1. Filtering

Filtering is a rule-based approach aimed to perform fast assessment of usefulness molecules in the given context. In drug design area, the filtering is used to eliminate compounds with unfavorable pharmacodynamic or pharmacokinetic properties as well as toxic compounds. Pharmacodynamics considers binding drug-like organic molecules (ligands) to chosen biological target. Since the efficiency of ligand-target interactions depends on spatial complementarity of their binding sites, the filtering is usually performed with 3D-pharmacophores, representing "optimal" spatial arrangements of steric and electronic features of ligands [83, 84]. Pharmacokinetics is mostly related to absorption, distribution, metabolism and excretion (ADME) related properties: octanol-water partition coefficients (*log P*), solubility in water (*log S*), blood-brain coefficient (*log BB*), partition coefficient between different tissues, skin penetration coefficient, etc.

Fragment descriptors are widely used for early ADME/Tox prediction both explicitly and implicitly. The easiest way to filter large databases concerns detecting undesirable molecular fragments (*structural alerts*). Appropriate lists of structural alerts are published for toxicity [85], mutagenicity [86], and carcinogenicity [87]. Klopman *et al* were the first to recognize the potency of using fragmental descriptors for this purpose [88-90]. Their programs CASE [88], MultiCASE [91, 92], as well as more recent MCASE QSAR expert systems [93], proved to be effective tools to assess mutagenicity [89, 92, 93] and carcinogenicity [90, 92] of organic compounds. In these programs, sets of biophores (analogs of structural alerts) were identified and used for activity predictions. A number of more sophisticated fragment-based expert systems of toxicity assessment - DEREK [94], TopKat [95] and Rex [96] − have been developed. DEREK is a knowledge-based system operating with human-coded or automatically generated [97] rules about toxicophores. Fragments in the DEREK knowledge base are defined by means of linear notation language PATRAN which codes the information about atom, bonds and stereochemistry. TopKat uses a large predefined set of fragment descriptors, whereas Rex implements a special kind of atom-pairs descriptors (*links*). To read more information about fragment-based computational assessment of toxicity, including mutagenicity and carcinogenicity, see review [98] and references therein.

The most popular filter used in drug design area is based on the Lipinsky "rule of five" [99], which takes into account the molecular weight, the number of hydrogen bond donors and acceptors, along with the octanol-water partition coefficient *logP*, to assess the bioavailability of oral drugs. Similar rules of "drug-likeness" or "lead-likeness" were later proposed by by Oprea [100], Veber [101] and Hann [102]. Formally, fragment descriptors are not explicitly involved there. However, many computational approaches to assess *logP* are fragment-based [51, 103, 104]; wheras H-donors and acceptor sites are simplest molecular fragments.

**3.2. Similarity search**

The similarity-based virtual screening is based on an assumption that all compounds in a chemical database, which are similar to a query compound, could also have similar biological activity. Although this hypothesis is not always valid (see discussion in [105]), quite often the set of retrieved compounds is enriched by actives [106].

To achieve high efficacy of similarity-based screening of databases containing millions compounds, molecular structures are usually represented by *screens* (structural keys) or fixed-size or variable-size *fingerprints*. Screens and fingerprints can contain both 2D- and 3D-information. However, the 2D-fingerprints, which are a kind of binary fragment descriptors, dominate in this area. Fragment-based structural keys, like MDL keys [48], are sufficiently good for handling small and medium-sized chemical databases, whereas processing of large databases is performed with fingerprints having much higher information density. Fragment-based Daylight [107], BCI [108] and UNITY 2D [109] fingerprints are the most known examples.

The most popular similarity measure for comparing chemical structures represented by means of fingerprints is the Tanimoto (or Jaccard) coefficient $T$ [110]. Two structures are usually considered similar if $T>0.85$ [106] (for Daylight fingerprints [107]). Using this threshold, Taylor estimated a probability to retrieve actives as 0.012-0.50 [111], whereas according to Delaney this number raises to 0.40-0.60 [112] (using Daylight fingerprints [107]). These computer experiments confirm usefulness of the similarity approach as an instrument of virtual screening.

Schneider *et al* have developed a special technique for performing virtual screening referred to as CATS (Chemically Advanced Template Search) [113]. In its framework chemical structures are described by means of so-called correlation vectors, each component of which equals to the number of times a certain atom pair is contained in a chemical structure divided by the total number of non-hydrogen atoms in it. Each atom in an atom pair is specified as belonging to one of five classes (hydrogen-bond donor, hydrogen-bond acceptor, positively charged, negatively charged, and lipophilic), while topological distances of up to 10 bonds are also considered in the atom-pair specification. The similarity between molecules is assessed in this approach using Euclidean distance between the corresponding correlation vectors. CATS was shown to outperform MERLIN with Daylight fingerprints [107] for retrieving thrombin inhibitors in a virtual screening experiment [113].

Hull *et al* have developed the *Latent Semantic Structure Indexing* (LaSSI) approach to perform similarity search in low-dimensional chemical space [114] [115]. To reduce the dimension of initial chemical space, the singular value decomposition method is applied for the descriptor-molecule matrix. Ranking molecules by similarity to a query molecule was performed in the reduced space using the cosine similarity measure [116], whereas the Carhart's atom pairs [68] and the Nilakantan's topological torsions [77] were used as descriptors. The authors claim that this approach "has several advantages over analogous ranking in the original descriptor space: matching latent structures is more robust than matching discrete descriptors, choosing the number of singular values provides a rational way to vary the 'fuzziness' of the search" [114].

The issue of "fuzzification" of similarity search was addressed by Horvath *et al* [70-72]. The first fuzzy similarity metric suggested in work [70] relies on partial similarity scores calculated with respect to the inter-atomic distances distributions for each pharmacophore pair. In this case the "fuzziness" enables to compare pairs of pharmacophores with different topological or 3D distances. Similar results [71] were achieved using fuzzy and weighted modified Dice similarity metric [116]. Fuzzy pharmacophore triplets FPT (see section 2) can be gradually mapped onto related basis triplets, thus minimizing binary classification artifacts [72]. In new similarity scoring index introduced in reference [72], the simultaneous absence of a pharmacophore triplet in two molecules is taken into account. However, this is a less-constraining indicator of similarity than simultaneous presence of triplets.

Most of similarity search approaches require only a single reference structure. However, in practice several compounds with the same type of biological activity are often available. This motivated Hert *et al* [117] to develop the *data fusion method* which allows one to screen a database using all available reference structures. Then, the similarity scores are combined for all retrieved structures using selected fusion rules. Searches conducted on the MDL Drug Data Report database using fragment-based UNITY 2D [109], BCI [108], and Daylight [107] fingerprints have proved the effectiveness of this approach.

The main drawback of the conventional similarity search concerns an inability to use experimental information on biological activity to adjust similarity measures. This results in inability to discriminate between relevant and non-relevant fragment descriptors being used for computing similarity measures. To tackle this problem, Cramer *et al* [34] developed *substructural analysis* in which each fragment (represented as a bit in a fingerprint) is weighted by taking into account its occurrence in active and in inactive compounds. Later on, many similar approaches have been described in the literature [118].

One-class classification [119] (or novelty detection [120, 121]) is a novel promising approach to conducting similarity-based virtual screening. It considers two types of compounds: active ("object class") and inactive ("outliers"). A new compound is predicted to be active if it lies in the dense area of the point cloud formed by active compounds contained in the training set and inactive if outside. Thus, this new compound is viewed as being similar to the set of active compounds. The fundamental difference between the one-class classification and the conventional similarity search is the ability of the former to use the whole training set instead of a single query compound in order to learn the optimal similarity measure for virtual screening. In accordance with this methodology, Karpov et al [69, 122] used the auto-associative neural networks and the one-class Support Vector Machines (1-SVM) in virtual screening (using fragment descriptors) against various biological targets.

One more way to conduct a similarity-based virtual screening is to retrieve the structures containing a user-defined set of "pharmacophoric" features. In *Dynamic Mapping of Consensus positions* (DMC) algorithm [123] those features are selected by finding common positions in bit strings for all active compounds. The *potency-scaled DMC* algorithm (POT-DMC) [124] is a modification of DMC in which compounds activities are taken into account. The latter two methods may be considered as intermediate between conventional similarity search and probabilistic SAR approaches.

Batista, Godden and Bajorath developed the MolBlaster method [125], in which molecular similarity is assessed by *Differential Shannon Entropy* [126] computed from populations of randomly generated fragments. For the range $0.64 < T < 0.99$, this similarity measure provides with the same ranking as the Tanimoto index *T*. However for the smaller values of *T* the entropy-based index is a more sensitive, since it distinguishes between pairs of molecules having almost identical *T*. To adapt this methodology for large-scale virtual screening, the *Proportional Shannon Entropy* (PSE) metrics was introduced [127]. A key feature of this approach is that class-specific PSE of random fragment distributions enables the identification of the molecules sharing with known active compounds a significant number of signature substructures.

Similarity search methods developed for individual compounds are difficult to apply directly for chemical reactions involving many species subdividing by two types: reactants and products. To overcome this problem, Varnek *et al* [15] suggested to condense all participating in reaction species in one only molecular graph (*Condensed Graphs of Reactions* (*CGR*) [15]) followed by its fragmentation and application of developed fingerprints in "classical" similarity search. Besides conventional chemical bonds (simple, double, aromatic, etc), a CGR contains dynamical bonds corresponding to created, broken or transformed bonds. This approach could be efficiently used for screening of large reaction databases.

It should be noted that the similarity concepts are widely used in selecting of diverse sets of compounds (see reviews [128-132] and references therein).

### 3.3. SAR/QSAR/QSPR models

Simplistic and heuristic similarity-based approaches can hardly produce as good predictive models as modern statistical and machine learning methods able to assess quantitatively biological or physicochemical properties [50]. QSAR-based virtual screening consists in direct assessment of activity values (numerical or binary) of all compounds in the database followed by selection of hits possessing desirable activity. Mathematical methods used for models preparation could be subdivided into *probabilistic* and *regression* approaches. The former assesses a probability that a given compound is active or not active whereas the latter numerically evaluate the activity values. A limited size of this paper doesn't allow us to cite all successful stories related to application of probabilistic and regression models in virtual screening; only some examples will be presented.

Harper *et al* [133] have demonstrated a much better performance of probabilistic *binary kernel discrimination* method to screen large databases compared to backpropagation neural networks or conventional similarity search. The Carhart's atom-pairs [68] and Nilakantan's topological torsions [77] were used as descriptors in that study.

Aiming to discover new cognition enhancers, Geronikaki *et al* [134] applied the PASS program [56], which implements a probabilistic Bayesian-based approach, and the DEREK rule-based system [94] to screen a database of highly diverse chemical compounds. Eight compounds with the highest probability of cognition-enhancing effect were selected. Experimental tests have shown that all of them possessed a pronounced antiamnesic effect.

Bender, Glen et al have applied [59-63] several probabilistic machine learning methods (naïve Bayesian classifier, inductive logic programming, and support vector inductive learning programming) in conjunction with circular fingerprints for making classification of bioactive chemical compounds and performing virtual screening on several biological targets. The latter of these three methods (*i.e.* support vector inductive learning programming) was shown to perform significantly better than the other two methods [63]. Advantages of using circular fingerprints were pointed out [61].

Regression QSAR/QSPR models are used to assess ADME/Tox properties or to detect "hit" molecules capable to bind a certain biological target. Available in the literature fragments based QSAR models for blood-brain barrier [135], skin permeation rate [136], blood-air [137] and tissue-air partition coefficients [137] could be mentioned as examples. Many theoretical approaches of calculation of octanol/water partition coefficient log *P* involve fragment descriptors. The methods by Rekker [138, 139], Leo and Hansch (CLOGP) [103, 140], Ghose-Crippen (ALOGP) [141-143], Wildman and Crippen [144], Suzuki and Kudo (CHEMICALC-2) [145], Convard (SMILOGP) [146], Wang (XLOGP) [147, 148] represent just a few modern examples. Fragment-based predictive models for estimation solubility in water [149] and DMSO [149] are available.

Benchmarking studies performed in references [135-137, 150] show that QSAR/QSPR models for various biological and physicochemical properties involving fragment descriptors are, at least, as robust as those involving topological, quantum, electrostatic and other types of descriptors.

### 3.4. *In silico* design

In this section we consider examples of virtual screening performed on a database containing only virtual (still non-synthesized or unavailable) compounds. Generation of virtual libraries is usually performed using combinatorial chemistry approaches [151-153]. One of simplest ways is to attach systematically user-defined substituents $R_1, R_2, ..., R_N$ to a given scaffold. If the list for the substituent $R_i$ contains $n_i$ candidates, the total number of generated structures is $N = \prod_i n_i$, although taking symmetry into account could reduce the library's size. The number of substituents $R_i$ ($n_i$) should be carefully selected in order to avoid a generation of too large set of structures (combinatorial explosion). The "optimal" substituents could be prepared using fragments selected at the QSAR stage, since their contributions into activity (for linear models) allow one to estimate an impact of combining the fragment into larger species ($R_i$). In such a way, a focused combinatorial library could be generated.

The technology based on combining QSAR, generation of virtual libraries and screening stages has been implemented into ISIDA program and applied to computer-aided design of new uranyl binders belonging to two different families of organic molecules: phosphoryl containing podands [154] and monoamides [155] . QSAR models have been developed using different machine-learning methods (multi-linear regression analysis, associative neural networks [156] and support vector machines [157]) and fragment descriptors (atom/bond sequences and augmented atoms). Then, these models were used to screen virtual combinatorial libraries containing up to 11000 compounds. Selected hits were synthesized and tested experimentally. Experimental data well correspond to predicted the uranyl binding affinity. Thus, initial data sets were significantly enriched with new efficient uranyl binders, and one of new molecules was found more efficient than previously studied compounds. A similar study was conducted for development of new 1-[2-hydroxyethoxy)methyl]-6-(phenylthio)thymine (HEPT) derivatives potentially possessing high anti-HIV activity [158]. This demonstrates universality of fragment descriptors and broad perspectives of their use in virtual screening and *in silico* design.

**Conclusion**

The power of fragment descriptors originates from their universality, very high computational efficacy, simplicity of interpretation, as well as their high diversity and versatility. The latest challenges in chemogenomics and high throughput virtual screening have raised their role in effective processing of huge amounts of relevant data and computer-aided design of new compounds.

**Acknowledgement**. The authors thank Dr I. Tetko and Dr G. Marcou for fruitful discussion.

**Disclosure**. This chapter is an updated version of the article [159].